# MoSAIC: Codon Harmonization of Monte Carlo-Based Simulated Annealing for Linked Codons in Heterologous Protein Expression

*Yoonho Jeong,[†] Chengcheng Yang,[‡] Ryan Fernandez Medina Hariri,[‡] Jihoo Kim,[†] Eok Kyun Lee,[†] Younghoon Lee,[†] Won June Kim,[§] Seung Seo Lee,[‡] and Insung S. Choi[*,†]*

[†]Department of Chemistry, KAIST, Daejeon 34141, Korea

[‡]School of Chemistry and Chemical Engineering, Highfield Campus, University of Southampton, Southampton SO17 1BJ, United Kingdom

[§]Department of Biology and Chemistry, Changwon National University, Changwon 51140, Korea

*Email: ischoi@kaist.ac.kr

ABSTRACT: Codon usage bias has a crucial impact on the translation efficiency and co-translational folding of proteins, necessitating the algorithmic development of codon optimization/harmonization methods, particularly for heterologous recombinant protein expression. Codon harmonization is especially valuable for proteins sensitive to translation rates, because it can potentially replicate native translation speeds, preserving proper folding and maintaining protein activity. This work proposes a Monte Carlo-based codon harmonization algorithm, MoSAiC (Monte Carlo-based Simulated Annealing for Linked Codons), for the harmonization of a set of linked codons, which differs from conventional codon harmonization, by focusing on the codon sets rather than individual ones. Our MoSAiC demonstrates robust computational performance on ribosomal proteins (S18, S15, S10, and L11) as model systems. Among them, the harmonized gene of RP S18 was expressed and compared with the expression of the wild-type gene. The harmonized gene clearly yielded a larger quantity of the protein, from which the amount of the soluble protein was also significant. These results underscored the potential of the linked codon harmonization approach to enhance the expression and functionality of sensitive proteins, setting the stage for more efficient production of recombinant proteins in various biotechnological and pharmaceutical applications.

## INTRODUCTION

Heterologous recombinant protein expression is a pivotal technique in synthetic biology and the biotechnology industry, used in the research and production of biotherapeutics, industrial enzymes, and vaccines. (1-4) A notable example is human insulin, the first genetically engineered recombinant protein manufactured by *Escherichia coli* (*E. coli*) and approved by the U.S. Food and Drug Administration (FDA) for therapeutic use. (5) Prior to the development of recombinant human insulin, insulin was primarily derived from animal sources or other humans, posing significant risks due to potential immunogenic risks and being costly and time-consuming. (6) The advent of recombinant human insulin has greatly propelled the research and development of other recombinant therapeutics, expanding therapeutic options and improving patient outcomes.

The genetic code, composed of triplet codons, is degenerate with 61 codons encoding 20 different amino acids. Except for methionine (Met, M) and tryptophan (Trp, W), each amino acid is represented by multiple synonymous codons, which are not used equally in organisms. This codon usage bias can significantly impact heterologous recombinant protein expression, as different host organisms, such as *E. coli*, exhibit preferences for synonymous codons, thereby influencing protein expression levels and folding efficiency. (7-10) For example, highly expressed genes tend to favor the codons that correspond to the most abundant tRNAs, enhancing the efficiency and speed of protein synthesis. Since the mRNA sequences with different synonymous codon usage can still translate into identical polypeptide sequences, careful genetic design is crucial to ensure that the produced protein is structurally and functionally equivalent to the target protein, maintaining its biological efficacy.

Codon optimization is a strategy employed primarily to enhance the heterologous recombinant protein expression by adjusting the codon sequence of genes to align with the preferred codon usage of the host. Originally, this was implemented using a ‘one amino acid–one codon’ approach, where rare codons are typically replaced with the ones that are more frequently used in the genome of the host organism. (11-14) Although this approach greatly increased protein expression levels—by up to $10^3$-fold, (15) it also disrupted the balance of the tRNA pool, consequently hindering the growth of the host organism. (11,16) In addition, proteins begin folding not only after translation is complete, but also concurrently with translation in a process known as co-translational folding. (17,18) Altering codons to more frequent ones changes the co-translational rhythm, potentially depriving nascent peptide chains of sufficient time to fold correctly, which can affect protein functions. (19-21) As a refinement of codon optimization, codon harmonization has recently been proposed to preserve the natural codon usage patterns of the source organism, while also considering the natural temporal sequence of protein folding. (4,22-24) Codon harmonization emulates the intrinsic translation rate and rhythm of the source organism, facilitating protein synthesis without co-translational misfolding. By aligning protein expression more closely with the cellular processes of the source organism, codon harmonization can improve protein solubility and functionality.

Recent years have witnessed significant developmental efforts in computer algorithms for codon optimization and harmonization. (25-28) As an early tool in codon optimization, EuGene operates in two stages: data-gathering and gene optimization. (25) During the optimization phase, EuGene restructures the gene sequence with six strategies: adjusting codon usage, refining codon context, managing GC content (G, guanine; C, cytosine), controlling hidden stop codons, eliminating repetitions, and removing deleterious sites. These adjustments are based on the information obtained in the data-gathering stage, which includes protein structures, orthologs, codon adaptation index (CAI), (29) relative synonymous codon usage, and codon pair bias. Within the open-source Galaxy platform, a specific tool can be utilized to align codon usage in synthetic genes with the natural frequency of codons in the target organism, aiming to reduce the codon harmonization index (CHI). (26) CHARMING, designed for codon harmonization, is grounded in translational kinetics and codon adjustment. (28) It features two metric modes (CHARMING:Geo and CHARMING:MM), which utilize CAI and $\%MinMax$, respectively, to estimate relative translation rates. These metrics help align the expression of synthetic genes more closely with the native protein synthesis processes of the source organism. CHARMING operates deterministically, following a predefined algorithm. However, since the parameters used to develop these algorithms, such as the termination criteria, are heuristic, it is unclear whether the optimized gene truly represents the best possible solution. Given the average of 3.05 synonymous codons per amino acid, the codon space for even a simple protein of 100 amino acids is vast, estimated at $10^{48}$ ($3.05^{100}$). With heuristic parameters guiding the optimization, there would be a risk that these algorithms may only achieve a local optimum in this immense codon space, rather than the global optimum. Deep learning methods such as bidirectional long-short-term memory conditional random fields and the T5 model have recently been applied to codon optimization and harmonization. (30-32)

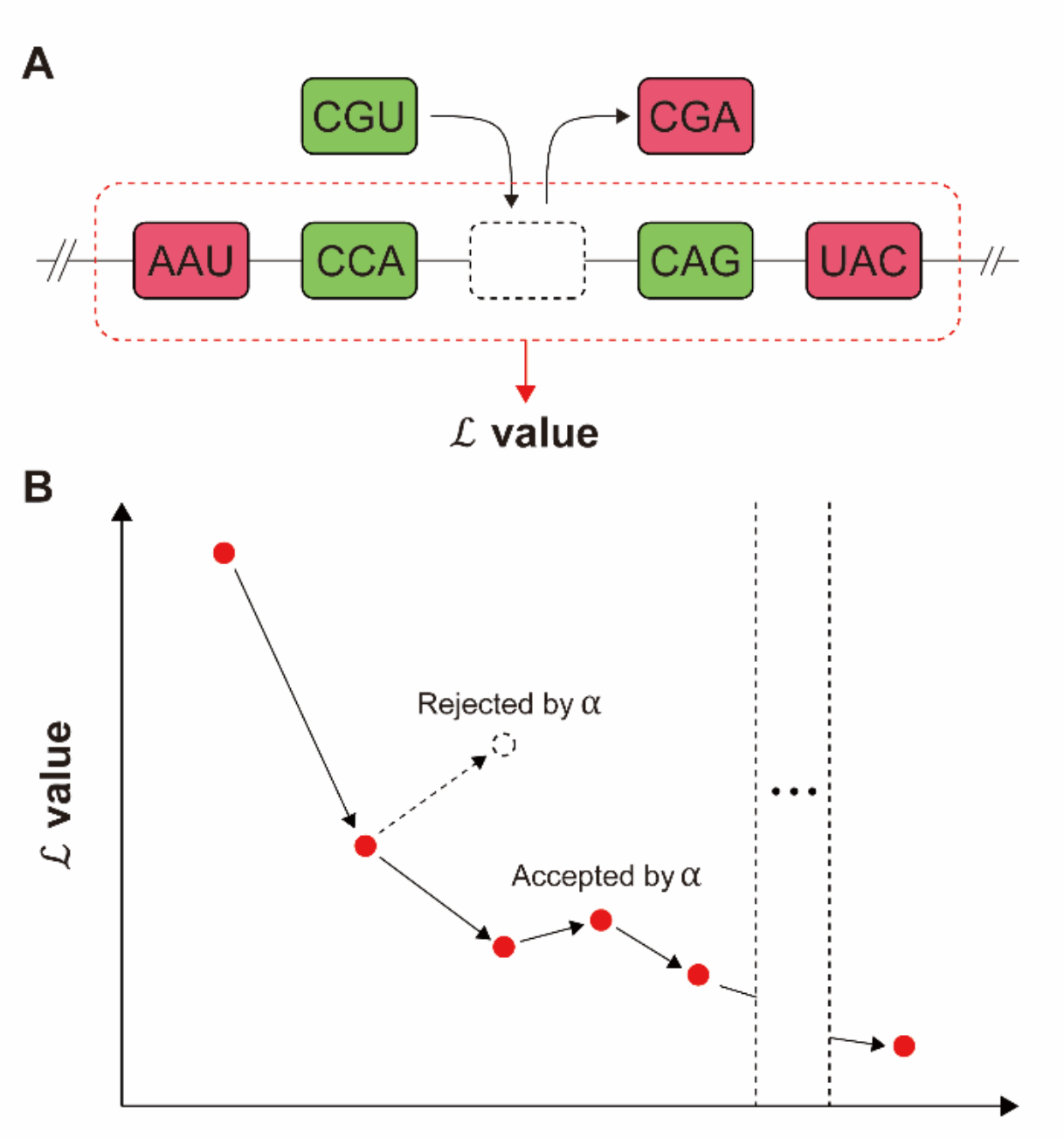

**Figure 1.** Overview of MoSAIC algorithm. (A) Joint optimization of synonymous codons within a sliding window, where each linked codon set is evaluated by the objective function $\mathcal{L}$. (B) Monte Carlo–based simulated annealing workflow for minimizing $\mathcal{L}$ through probabilistic acceptance $\boldsymbol{\alpha}$.

In this paper, we propose a simple yet versatile algorithm designed to generate codon sequences that closely mirror the translation rhythm of the source organism. Our method systematically replaces codons in the original gene sequence, derived from *Homo sapiens* (*H. sapiens*) in this study, with those conforming to the frequency distribution observed in a target host organism, specifically *E. coli*. Unlike the conventional methods that focus on aligning individual codon usage based on the frequency order, our algorithm prioritizes the harmonization of linked codons, a process we term Monte Carlo-based Simulated Annealing for Linked Codons (MoSAiC) (Figure 1A). The traditional approaches face limitations in achieving optimal substitution due to the inherent differences in codon usage biases among organisms. For instance, while CCC and CCG are the most frequently used codons for proline (Pro or P) within the synonymous set of CCC, CCG, CCU, and CCA (U: uracil; A: adenine) in *H. sapiens* and *E. coli*, respectively, direct replacement of CCC with CCG does not account for the normalized frequency discrepancy: 0.33 in *H. sapiens* vs. 0.49 in *E. coli*. This mismatch indicates that codon matching, based solely on frequency order at the single codon level, can lead to suboptimal outcomes. It is envisaged that harmonization of frequency patterns across multiple consecutive codons can significantly enhance the effectiveness of codon harmonization strategies.

## RESULTS AND DISCUSSION

**MoSAiC Algorithm.** We adopted the previously proposed $\%MinMax$ metric to analyze codon usage patterns. (33,34) In brief, the contiguous segments in a codon sequence are grouped into a single window, and the $\%MinMax$ values of the grouped codons are calculated. The choice of an appropriate window size $z$ is critical. When the window size is too small, the resulting $\%MinMax$ profiles become excessively noisy and highly sensitive to local fluctuations. Conversely, larger window sizes tend to yield lower loss values, but this apparent improvement is accompanied by excessive profile smoothing and reduced codon-level resolution. As shown in Figure S1, $z = 10$ provided a practical balance between noise reduction and signal preservation, avoiding both excessive fluctuations at smaller window sizes and over-smoothing at larger ones. In addition, this choice is biologically relevant, as a ribosome typically spans 28-30 nucleotides during translation, corresponding to approximately 10 codons.

$$\bar{f}_{i\cdot(\Sigma)} = \frac{1}{n_i}\sum_{j=1}^{n_i} f_{ij(\Sigma)} \qquad \text{(Equation 1)}$$

$$\Delta f = \sum_{i=1}^{z}\left(f_{i(\Sigma)}^{*} - \bar{f}_{i\cdot(\Sigma)}\right) \qquad \text{(Equation 2)}$$

$$\mathcal{M}_{(\Sigma)} = \begin{cases} \dfrac{\sum_{i=1}^{z}\left(f_{i(\Sigma)}^{*} - \bar{f}_{i\cdot(\Sigma)}\right)}{\sum_{i=1}^{z}\left(f_{i(\Sigma)}^{M} - \bar{f}_{i\cdot(\Sigma)}\right)} & \Delta f \geq 0 \\ -\dfrac{\sum_{i=1}^{z}\left(\bar{f}_{i\cdot(\Sigma)} - f_{i(\Sigma)}^{*}\right)}{\sum_{i=1}^{z}\left(\bar{f}_{i\cdot(\Sigma)} - f_{i(\Sigma)}^{m}\right)} & \Delta f < 0 \end{cases} \qquad \text{(Equation 3)}$$

In Equation 1, $\bar{f}_{i\cdot(\Sigma)}$ is the average frequency of the synonymous codons for the *i*th amino acid in the species, $\Sigma$, from which the sequence derives ($H$ for *H. sapiens* and $E$ for *E. coli* in this paper). $f_{ij(\Sigma)}$ is the frequency of the *j*th synonymous codon for the *i*th amino acid in the species $\Sigma$, and $n_i$ is the number of synonymous codons. If the sum of frequencies of observed codons ($f_{i(\Sigma)}^{*}$'s) exceeds the sum of $\bar{f}_{i\cdot(\Sigma)}$'s in a given window, a positive $\%Max_{(\Sigma)}$ value is calculated, where $f_{i(\Sigma)}^{M}$ is the frequency of the most frequent codon among the synonymous codons for the *i*th amino acid in the $\Sigma$ species. Likewise, the negative $\%Min_{(\Sigma)}$ value is calculated, where $f_{i(\Sigma)}^{m}$ is the frequency of the rarest codon. A $\%MinMax_{(\Sigma)}$ pattern is constructed as the window slides along the sequence with a stride of one. Zero padding is applied to both sides of the $\%MinMax_{(\Sigma)}$ pattern for length consistency, leading to the generation of a $\%MinMax_{(\Sigma)}$ profile for a sequence $\boldsymbol{x}^*$, denoted as $\mathcal{M}_{(\Sigma)}(\boldsymbol{x}^*)$ (Equation 2, 3).

For linked codon harmonization, MoSAIC explores the search space widely at the beginning and gradually focuses on local improvements. (35) Figure 1B depicts the workflow of our MoSAIC algorithm. Specifically, the initiation involves the generation of an initial sequence, denoted as $\boldsymbol{x}^{(0)}$, for a good starting point. The sequence is formed by replacing the codons in a wild-type (WT) sequence, $\boldsymbol{x}^{WT}$, with corresponding codons from *E. coli*. The codon initialization is based on matching the codon frequency order between *H. sapiens* and *E. coli*. Formally, $\boldsymbol{x}^{(0)}$ is obtained using the conventional codon harmonization method. (4) In each iteration of the MoSAIC algorithm, a test sequence, $\widetilde{\boldsymbol{x}}^{(t)}$, is generated from $\boldsymbol{x}^{(t)}$ $(t = 0, 1, 2, \cdots)$ by randomly substituting one codon in the sequence with a synonymous codon.

A loss ($\mathcal{L}$) is formally defined as the sum of absolute differences between the $\%MinMax$ values of $\mathcal{M}_{(H)}(\boldsymbol{x}^{WT})$ and $\mathcal{M}_{(E)}\left(\boldsymbol{x}^{(t)}\right)$ (Equation 4), where $W$ denotes the length of a $\%MinMax$ profile. In the present study, the loss function is interpreted as a measure of similarity in $\%MinMax$ profiles rather than as a direct predictor of protein expression level. The algorithm iteratively computes $\mathcal{L}$ until an optimally harmonized sequence is found. The number of iterations is proportional to the sequence length, because longer sequences require more opportunities for MoSAIC.

$$\mathcal{L}(\boldsymbol{x}^{(t)}) = \sum_{k=1}^{W} \left| \mathcal{M}_{(H)}(\boldsymbol{x}^{WT})_k - \mathcal{M}_{(E)}\left(\boldsymbol{x}^{(t)}\right)_k \right| \quad \text{(Equation 4)}$$

$$\alpha = \exp\left(\frac{\mathcal{L}(\boldsymbol{x}^{(t)}) - \mathcal{L}(\widetilde{\boldsymbol{x}}^{(t)})}{\tau^{(t)}}\right) \quad \text{(Equation 5)}$$

$$\tau^{(t)} = \tau^{(0)} - \eta t \quad \text{(Equation 6)}$$

In each MoSAIC iteration, the algorithm computes the acceptance probability $\alpha$ from the difference between $\mathcal{L}(\boldsymbol{x}^{(t)})$ and $\mathcal{L}(\widetilde{\boldsymbol{x}}^{(t)})$, with a simulated annealing formula (Equation 5). In this work, the acceptance rate was modulated by introducing a temperature parameter, $\tau$, to the equation for $\alpha$, and a linear cooling schedule was employed to ensure legitimate acceptance chances with steady decrease (Equation 6). We empirically evaluated alternative cooling strategies, including exponential and logarithmic schedules, within the same optimization framework. In our setting, these schemes caused $\tau^{(t)}$ to decay too rapidly during the early stages of optimization, which frequently led to premature convergence to local minima. The decreasing rate, $\eta$, is set to be the initial $\tau^{(0)}$ value divided by the total number of iterations, with $\tau^{(t)}$ approaching zero as the iteration progresses. Therefore, an appropriate initial temperature and the total number of iterations are necessary, as the decreasing rate should be sufficiently low to ensure that the probability distribution can reach thermodynamic equilibrium throughout the optimization process. In the case that $\mathcal{L}\left(\widetilde{\boldsymbol{x}}^{(t)}\right)$ is greater than $\mathcal{L}\left(\boldsymbol{x}^{(t)}\right)$, a large $\tau^{(t)}$ increases the acceptance chance for $\widetilde{\boldsymbol{x}}^{(t)}$, and vice versa.

If $\min(\alpha, 1)$ is bigger than $u$, randomly sampled from a uniform distribution $\mathrm{U}(0, 1)$, $\widetilde{\boldsymbol{x}}^{(t)}$ is accepted as the harmonized sequence and becomes the target $\boldsymbol{x}^{(t+1)}$ for the next iteration.

Otherwise, $\tilde{\boldsymbol{x}}^{(t)}$ is rejected and the previous sequence $\boldsymbol{x}^{(t)}$ is passed to the next iteration. This stochastic acceptance is employed to explore a wider codon space, increasing the likelihood to reach the global minimum.

**Model Proteins: Ribosomal Proteins.** We selected ribosomal proteins (RPs) as a model system in this study. The ribosome performs the essential function of synthesizing proteins within all living cells. (36,37) Eukaryotic ribosomes consist of four ribosomal RNAs and approximately 80 RPs (33 in the small 40S subunit and about 47 in the large 60S subunit for *H. sapiens*). (38)

In general, proper protein folding occurs naturally or is assisted by molecular chaperones that facilitate the folding process. However, in heterologous recombinant protein expression, the gene is foreign to the host organism and the intracellular environment differs from the native context. These factors often result in the production of misfolded proteins that aggregate into insoluble and inactive inclusion bodies. RPs, in particular, contain a high proportion of positively charged, basic amino acids, such as arginine (Arg, R) and lysine (Lys, K), which can cause aggregation during heterologous recombinant expression. Correspondingly, the recovery of RPs from inclusion bodies remains challenging, and some attempts have been made to facilitate refolding by adding fusion tags like glutathione S-transferase (GST) tag, maltose-binding protein (MBP) tag, and histidine tag (His-tag). (39-41) Among RPs, we selected S18, S15, S10, and L11, because they share similar sizes, isoelectric points, and amino acid compositions.

**Performance: Codon Harmonization.** In the simulated annealing process, the initial temperature, $\tau^{(0)}$, not only influences acceptance probability but also regulates the decreasing rate. For example, if $\tau^{(0)}$ is set too high, the computational cost increases, and the utility of the sequence initialization diminishes due to unnecessary acceptance at the beginning. We, therefore, systematically screened and optimized the $\tau^{(0)}$ value for each protein, prior to the experiments. The initial temperature parameter $\tau^{(0)}$ was selected empirically by testing a range of candidate values and choosing the value that produced the lowest loss. Because the simulated annealing procedure is stochastic, the influence of $\tau^{(0)}$ on the optimization process cannot be characterized deterministically, and the current approach should therefore be regarded as a practical heuristic for parameter selection. Specifically, initial $\tau^{(0)}$ values were set to range from 1 to 17 in increments of 2, and computational results yielded lower loss values for $\tau^{(0)}$ values between 11 and 15 (Table S1A). Further refinement with 0.5 increments identified the optimal $\tau^{(0)}$ values of 14, 13, 13, and 11.5 for RP S18, S15, S10, and L11, respectively (Table S1B). On the other hand, since MOSAIC randomly changes one codon at each iteration, it inherently requires a number of iterations proportional to the length of the amino acid sequence. In this study, the maximum number of iterations was defined as 10,000 times the sequence length. During MOSAIC iterations, $\boldsymbol{x}^{(t)}$ was updated, and the $\%MinMax$ profile of the harmonized sequence, $\mathcal{M}_{(E)}\left(\boldsymbol{x}^{(T)}\right)$, was obtained.

We showcased our MOSAIC with RP S18, composed of 152 amino acids, as an example. Among the four RPs studied, the WT sequences of *H. sapiens* RP S15 and S18 tended to form inclusion bodies in *E. coli*, while S10 and L11 have been successfully expressed in soluble form. (39) Furthermore, although efforts have been made to enhance the solubility of S15 and S18 using a thioredoxin fusion protein, the results showed in little to no improvement for S18.

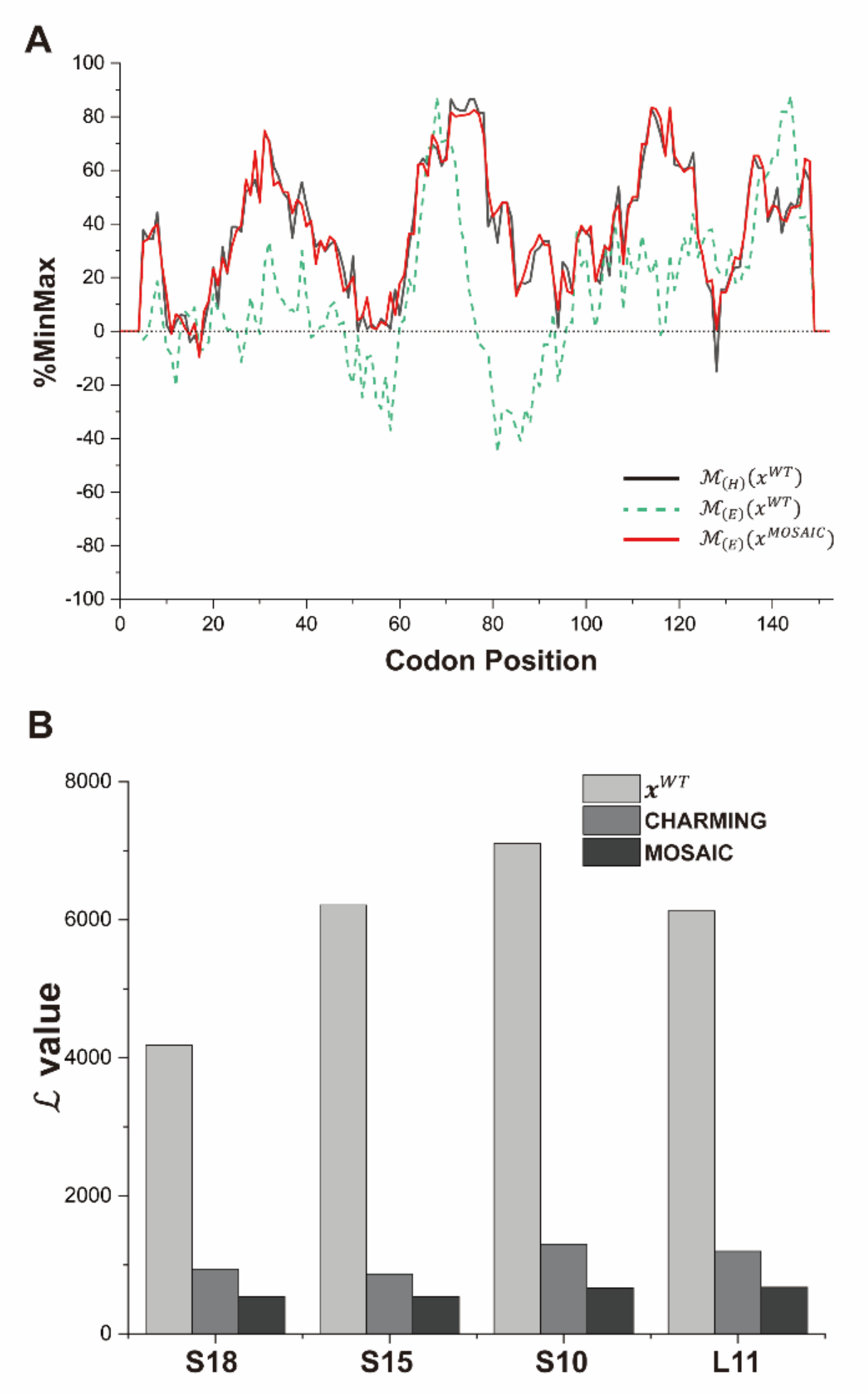

**Figure 2.** (A) $\%\boldsymbol{MinMax}$ profiles of RP S18. Solid black: a reference, $\boldsymbol{\mathcal{M}}_{(\boldsymbol{H})}(\boldsymbol{x}^{\boldsymbol{WT}})$; solid red: MOSAIC result, $\boldsymbol{\mathcal{M}}_{(\boldsymbol{E})}(\boldsymbol{x}^{(\boldsymbol{T})})$; dashed green: $\boldsymbol{\mathcal{M}}_{(\boldsymbol{E})}(\boldsymbol{x}^{\boldsymbol{WT}})$. (B) Loss ($\boldsymbol{\mathcal{L}}$) values of the model proteins for $\boldsymbol{x}^{\boldsymbol{WT}}$, CHARMING and MOSAIC results.

Predictably, the $\%MinMax$ profile of the WT sequences of *H. sapiens* RP S18, $\mathcal{M}_{(E)}(\boldsymbol{x}^{WT})$, computed using the codon usage of *E. coli*, differed markedly from that of $\mathcal{M}_{(H)}(\boldsymbol{x}^{WT})$, due to inherent codon bias (Figure 2A, solid black line versus dashed green line). Correspondingly, the initial $\mathcal{L}$ value of $\mathcal{M}_{(E)}(\boldsymbol{x}^{WT})$, referenced against $\mathcal{M}_{(H)}(\boldsymbol{x}^{WT})$, was calculated to be 4180.6. This value decreased following sequence initialization to $\boldsymbol{x}^{(0)}$, bringing it closer to $\mathcal{M}_{(H)}(\boldsymbol{x}^{WT})$ by a conventional codon harmonization method. (4) However, the $\mathcal{M}_{(E)}\left(\boldsymbol{x}^{(0)}\right)$ profile of RP S18 lay below the $\mathcal{M}_{(H)}(\boldsymbol{x}^{WT})$ profile, suggesting that many, if not all, segments were replaced with rare codons in the process of sequence initialization. This phenomenon was similarly observed in other model proteins. To verify this hypothesis, we screened 20,070 protein-coding sequences extracted from 58,506 genes of T2T-CHM13, the reference genome of *H. sapiens*. (42) Among these, 19,780 sequences exhibited the same pattern in the $\%MinMax$ plots after sequence initialization to $\boldsymbol{x}^{(0)}$ (Figure S2). The analysis indicated that simple codon substitution solely based on the codon usage frequency tends to increase the occurrence of rare codons in heterologous protein expression. As aforementioned, increasing the usage of rare codons is inefficient, because it can decrease translational efficiency and cause improper folding in those regions.

MOSAIC showed a notable performance for RP S18 and significantly reduced the loss to an average of 536.4 over 10 trials with a minimum value of 519.2 (solid red line in Figure 2A and Table S2). Although MOSAIC is a stochastic algorithm, the standard deviation across 10 trials was only 7.6, indicating that the algorithm is robust in harmonization performance. In comparison, CHARMING, (28) one of the best codon harmonization tools, decreased the loss to 938.9. Although different codon-optimization methods may be developed using different internal objective functions, the comparison in Figure 2 was performed using a common loss function applied uniformly to all resulting sequences. The reported loss values are therefore directly comparable across methods. We also calculated the harmonization percentage, which is defined as the percentage reduction in the loss relative to the initial loss of 4180.6. MOSAIC harmonized the WT sequence by 87.6%, while CHARMING achieved 77.5%. Furthermore, MOSAIC more accurately reproduced the sharp peaks in the $\%MinMax$ profile than CHARMING (Figure S3). Since the sites where translation shifts occur may play a pivotal role in protein synthesis, precise alignment of these regions may be critical for optimizing recombinant protein production. (10,43-44) Moreover, multiple features influence the translation process during protein synthesis, with the GC content of mRNA being one of the critical determinants of mRNA stability. mRNAs with excessively low GC content may lack sufficient hydrogen-bonding interactions to maintain stability. Conversely, excessively high GC content can hinder unwinding during transcription and translation, leading to ribosome stalling and reduced translation efficiency. The GC content of the MoSAIC-harmonized RP S18 sequence was 48.4%, which was close to that of the WT sequence (52.5%). Maintaining GC content within an intermediate range may help balance mRNA stability and translation initiation efficiency by preventing the formation of excessively stable mRNA secondary structures. Consistent with this, all three sequences fall within a commonly reported GC-content range in previous studies. (45-47)

Furthermore, we used RNAfold, a software package for mRNA secondary structure prediction included in the ViennaRNA Package 2.0, (48) to predict the secondary structure of each sequence. RNAfold calculates the minimum free energy (MFE) structures of RNA sequences and evaluates their thermodynamic stability and base-pairing probabilities. Notably, the MoSAIC-harmonized RP S18 sequence was predicted to have a more stable centroid secondary structure compared with other harmonized sequences, even though MoSAIC does not explicitly account for mRNA stability. The predicted MFE of the WT sequence was -128.6 kcal/mol, while those of the initialized sequence, CHARMING, and MoSAIC were -93.6, -73.8, and -114.6 kcal/mol, respectively. Taken together, the analyses showed that the sequence harmonized by MoSAIC, $\boldsymbol{x}^{(T)}$, closely mirrored the translation rhythm of $\boldsymbol{x}^{WT}$ in *H. sapiens*, and was expected to enhance translation efficiency and protein solubility.

The best performance of MoSAIC was also observed for three other model proteins, S15, S10, and L11 (Figure 2B and Table S2). For instance, in the case of RP S15, which forms insoluble aggregates when heterologously expressed in *E. coli*, the initial $\mathcal{L}$ value of 6212.4 was reduced to an average of 536.7 with MoSAIC, compared with 867.0 obtained using CHARMING. Similarly, the average $\mathcal{L}$ values of RP S10 and RP L11 decreased to 663.8 and 675.9 from 7100.3 and 6127.4, respectively, after MoSAIC-based harmonization. Overall, the computational results demonstrated that MoSAIC performs consistently across diverse target proteins while preserving native $\%MinMax$ profile characteristics in the generated coding sequences.

**Experimental Validation: RP S18.** To demonstrate the effectiveness of codon harmonization by MoSAIC, RP S18 was selected as the model protein for heterologous expression in *E. coli*. A harmonized gene (hRPS18) was cloned into the pET28a vector, and a laboratory *E. coli* strain BL21(DE3) pLysS was used as an expression host. In this study, experimental validation was limited to comparisons between MoSAIC-harmonized sequence and the wild-type gene. The protein was expressed under the induction of isopropyl β-D-thiogalactopyranoside (IPTG). As a control, the wild-type RP S18 gene (wtRPS18) was cloned into the same vector and expressed under identical conditions. The IPTG induction was performed with 1 mM IPTG and at 25 °C.

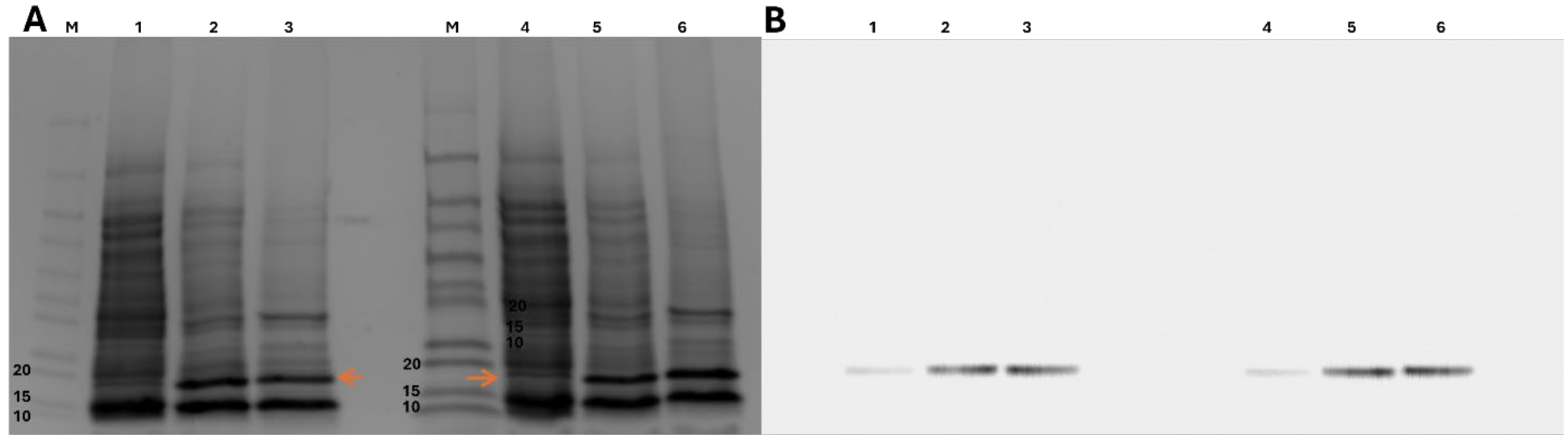

**Figure 3.** A representative heterologous expression of RP S18. (A) SDS PAGE with Coomassie Blue staining. Lane 1: soluble fraction of lysed BL21(DE3) pLysS harboring the harmonized gene (hRPS18); lane 2: bacterial lysate of BL21(DE3) pLysS harboring hRPS18; lane 3: inclusion body from lysed BL21(DE3) pLysS harboring the hRPS18; lane 4: soluble fraction of lysed BL21(DE3) pLysS harboring the wild-type gene (wtRPS18); lane 5: bacterial lysate of BL21(DE3) pLysS harboring wtRPS18, lane 6: inclusion body from BL21(DE3) pLysS harboring wtRPS18; M: protein molecular weight markers with three annotated as 10, 15, and 20 kDa, respectively. RP S18 is marked with a red arrow. (B) The Western blot of the SDS-PAGE gel. Coomassie Blue staining of the SDS-PAGE gel of (A) shows equal protein loading across lanes corresponding to the Western blot.

Sodium dodecyl sulfate–polyacrylamide gel electrophoresis (SDS-PAGE) analysis was performed by loading equal amounts of protein from each sample (Figure 3). The Western blot analysis showed that both the hRPS18 protein and the wtRPS18 protein, with an estimated molecular weight of 17.7 kDa, were expressed and detected in both the soluble fraction and the inclusion body (Figure 3B). To assess the differences in relative expression levels of the target protein between the hRPS18 and wtRPS18 constructs, densitometric analysis of Coomassie-stained SDS-PAGE gels was performed using ImageJ. (4,49) Integrated density values were obtained for the entire lanes of both constructs, and the integrated density of the target protein band was expressed as a percentage of the total lane integrated density. The analysis was conducted across four independent biological replicates, and the raw integrated density values for each lane and target band are provided in Supplementary Table 3 (Table S3). For each biological replicate, a fold change (FC) was calculated by dividing the relative band intensity of the hRPS18 construct by that of the wtRPS18 construct (Table 1 and Table S3). FC values greater than 1 indicate higher proportional expression in the hRPS18 construct. These analyses clearly showed that the hRPS18 construct consistently yielded a greater amount of soluble target protein than the wtRPS18 construct. On average, the harmonized construct exhibited approximately 30% higher soluble RP S18 expression than the wtRPS18 construct. Similarly, in the bacterial lysate, the hRPS18 construct showed approximately 20% higher than the wtRPS18 construct. Although the expression level varied substantially across biological replicates (Table S3), the consistently greater-than-1 FC values for soluble protein expression in hRPS18 provide supportive evidence that MoSAIC promotes more favorable protein expression in a heterologous host.

**Table 1.** Fold changes (FCs) in relative RP S18 expression between hRPS18 and wtRPS18 constructs across three sample fractions. Data are presented as mean ± SEM ($n = 4$).

| Sample | FC |
| --- | --- |
| soluble fraction | 1.327 ± 0.195 |
| bacterial lysate | 1.192 ± 0.215 |
| inclusion body | 2.371 ± 0.928 |

## CONCLUSIONS

In this work, we developed an effective codon harmonization algorithm, MoSAIC, which harmonizes the codon usage patterns of linked codons with the source organism, with convergence close to the global minimum within the vast codon space. The performance of MOSAIC was demonstrated through the heterologous expression of RP S18 in the *E. coli* strain, BL21(DE3) pLysS. Compared with the wild-type gene, the harmonized gene produced a respectably higher total yield and a greater soluble fraction of RP S18, demonstrating the robustness of the algorithm. We note that the current experimental validation is limited to a single representative protein and therefore should be interpreted as proof-of-concept support for the MOSAIC framework rather than as a comprehensive demonstration of its general applicability across diverse targets. Although the computational analyses indicate robust performance across multiple proteins, broader experimental validation will be necessary to determine the extent to which these benefits are consistently observed in wet-lab settings. Systematic multi-protein benchmarking will therefore be an important direction for future work.

Although the performance of MOSAIC was tested only using RPs as the model proteins, MOSAIC relies solely on sequence and codon usage information for linked codon harmonization. This feature makes it broadly applicable to various protein types, including membrane and secretory proteins. In addition to its robust computational performance for RPs, MOSAIC algorithm offers several advantages, including scalability, which allows for the incorporation of additional parameters into the linked codon harmonization process—such as GC content, mRNA secondary structures, and translation-related features—enabling highly customized sequence designs for target proteins. Further studies will evaluate the expression level and solubility of proteins optimized by MOSAIC. The resulting experimental data will help classify protein types that strongly depend on translation rhythm and identify regions requiring critical translational pauses for proper folding, providing insights both for fundamental studies and for further algorithmic optimization.

## DATA AND METHODS

**Data Preparation.** The gene sequence information for RPs was obtained from the Consensus CDS (CCDS) database, (50) specifically CCDS4771.1, CCDS12067.1, CCDS4792.1, and CCDS238.1 for RP S18, S15, S10, and L11, respectively. Codon usage frequencies used to calculate $\%MinMax$ values were obtained from the Codon and Codon-Pair Usage Tables (CoCoPUTs). (51) CoCoPUTs provides comprehensive codon, codon-pair, and dinucleotide usage data for all species represented in GenBank and RefSeq. (52,53)

To assess whether the sequence initialization of human coding sequences exhibits a preference for substitution toward rare codons, the telomere-to-telomere (T2T-CHM13) genome assembly was obtained from RefSeq. The T2T-CHM13 assembly represents the first complete human reference genome, assembled without gaps. From 58,506 annotated reference sequences, 20,070 coding sequences (CDSs) were extracted using the RefSeq annotation. These CDSs were initialized using the conventional codon harmonization method. The averages of $\%MinMax$ values of the initialized sequences were calculated, and sequences were counted if their averages were lower than those of the corresponding original sequences.

**Experimental Validation.** All the reagents were sterilized either by filtration through a 0.2 μm syringe filter or by autoclaving. Plasmids harboring a harmonized RP S18 gene (hRPS18) and the wild-type RP S18 gene (wtRPS18) on the pET28a vector were purchased from GenScript. Each plasmid, dissolved in sterilized deionized water, was introduced into the laboratory *E. coli* strain BL21(DE3) pLysS (Merck Life Science) using the heat shock method. Transformed cells were mixed with SOC media and incubated at 250 rpm and 37 °C for 2 h before plating on a Lysogeny broth (LB) agar plate (Fisher Scientific, UK) supplemented with kanamycin and chloramphenicol, which was incubated overnight at 37 °C. A few grown colonies were picked and suspended in the LB media supplemented with kanamycin and chloramphenicol, which was incubated in a shaking incubator at 250 rpm and 37 °C overnight. 10 mL of the overnight culture was added to 1 L of LB media supplemented with kanamycin and chloramphenicol, and incubated at 37 °C in a shaking incubator (250 rpm) until the optical density at 600 nm ($OD_{600}$) reached 0.5–0.7. At that point, IPTG was added to a final concentration of 1 mM, and the cells were further incubated at 25 °C for 16 h.

Cultured cells were centrifuged at 8,000 rpm and the supernatant was discarded. Cell pellets were resuspended in a binding buffer (pH 7.8, 20 mM Tris-HCl, 500 mM NaCl, 5 mM imidazole) supplemented with DNase I, lysozyme and protease inhibitors. The mixture was stirred on ice for 20 min and then sonicated on ice for lysis. The lysed cell suspension was centrifuged at 20,000 rpm for 35 min, and the supernatant and pellet were collected for analysis. Before SDS-PAGE was performed, the protein concentrations of all samples were measured using the Bradford assay and normalized to equal levels. SDS-PAGE was performed using the Mini-PROTEAN system with 4–15% precast gels. The gel was stained with Coomassie Brilliant Blue and the gel images were acquired using a GelDoc™ EZ Imager (Bio-Rad).

**Image Analysis.** Densitometric analysis was performed using ImageJ. For each gel image, lane profiles were generated by defining rectangular regions of interests (ROIs) over each lane, starting and ending at the first and last band respectively. Integrated densities were calculated for each lane. The integrated density of the band corresponding to the protein of interest (approximately 18 kDa, confirmed by co-migration with a molecular weight ladder and the Western blot) was expressed as a proportion of the total lane intensity, yielding a measured percentage of target protein abundance relative to total soluble or insoluble protein content. This normalization approach allows for lane-to-lane comparison, normalizing for variation in sample preparation. Analysis was performed across four independent biological replicates. For each replicate, a fold change was calculated by dividing the relative band intensity of the harmonised construct by that of the original construct for each matched fraction (bacterial lysate, soluble fraction, and inclusion body). The mean fold change and standard error were calculated across all four replicates with the numerical spreadsheet available in the supplementary information.

**Western Blot.** Proteins were transferred to a PVDF membrane using the iBlot™ 2 Gel Transfer Device (Thermo Fisher Life Sciences). The membrane was then blocked with 3% BSA in TBST to prevent non-specific binding. Following blocking, the membrane was incubated with HRP-conjugated HIS.H8 antibody diluted 1:5000 in blocking buffer for 2 h at room temperature. After antibody incubation, the membrane was washed three times with TBST to remove unbound antibody. Immunoreactive bands were detected using Cytiva ECL reagent prepared at a 1:1 ratio and visualized using a G:BOX (Syngene) imaging system.

**Availability and Implementation**

MOSAIC is freely available on https://github.com/CIS-group/MOSAIC.

ASSOCIATED CONTENT

**Supporting Information**.

The Supporting Information is available free of charge.

Effect of window size on loss and $\%MinMax$ profiles. (A) Normalized loss values for six proteins (FALVAC-1, PTP4A3, PA, PAE, Mmpl3, and HPDF) as a function of window size. For all proteins, the loss decreases monotonically with increasing window size, indicating improved smoothing and agreement with the target profile at larger window sizes. (B) $\%MinMax$ profiles of the RP S10 protein generated using different window sizes, arranged from left to right, top to bottom (window sizes: 5, 10, 15, 20, 25, and 30); Distribution of total $\%MinMax$ profile differences between original and initialized coding sequences. For each sequence, the total difference was calculated as the mean difference across the $\%MinMax$ profiles. CDS: coding sequence; $\%MinMax$ profiles. (A) RP S18, (B) RP S15, (C) S10, and (D) L11; The loss ($\mathcal{L}$) values of the experiments for optimizing $\tau^{(0)}$ values for the proteins. During the experiments, the maximum number of iterations was set to 1,000 times the sequence length. (A) Loss ($\mathcal{L}$) values for RP S18, S15, S10, and L11 from 1 to 17 of initial temperature by 2. (B) Loss ($\mathcal{L}$) values for RP S18, S15, S10, and L11 from 11 to 15 of initial temperature by 0.5; Loss ($L$) values for RP S18, S15, S10, and L11 at each optimal initial temperature, respectively; Raw Integrated Density (RID) values obtained by densitometric analysis in ImageJ for each lane and target protein band across four biological replicates. Relative band intensity (% of Total) was calculated by dividing target band raw integrated density by the corresponding construct total lane raw integrated density and multiplying by 100. (PDF)

AUTHOR INFORMATION

**Corresponding Author**

**Insung S. Choi** – Department of Chemistry, KAIST, Daejeon 34141, Korea

orcid.org/0000-0002-9546-673X;

Email: ischoi@kaist.ac.kr

**Author**

**Yoonho Jeong** – Department of Chemistry, KAIST, Daejeon 34141, Korea

**Chengcheng Yang** – School of Chemistry and Chemical Engineering, Highfield Campus, University of Southampton, Southampton SO17 1BJ, United Kingdom

**Ryan Fernandez Medina Hariri** – School of Chemistry and Chemical Engineering, Highfield Campus, University of Southampton, Southampton SO17 1BJ, United Kingdom; orcid.org/0009-0006-9865-6252

**Jihoo Kim** – Department of Chemistry, KAIST, Daejeon 34141, Korea

**Eok Kyun Lee** – Department of Chemistry, KAIST, Daejeon 34141, Korea

**Younghoon Lee** – Department of Chemistry, KAIST, Daejeon 34141, Korea

**Won June Kim** – Department of Biology and Chemistry, Changwon National University, Changwon 51140, Korea

**Seung Seo Lee** – School of Chemistry and Chemical Engineering, Highfield Campus, University of Southampton, Southampton SO17 1BJ, United Kingdom; orcid.org/0000-0002-8598-3303

## Author Contributions

I.S.C., Y.J., J.K., E.K.L, Y.L., and W.J.K. initiated the project. I.S.C., E.K.L., Y.L., and S.S.L. supervised the project. Y.L. and Y.J. conceptualized, Y.J., J.K., and E.K.L developed the codon harmonization algorithm. Y.J. performed the codon harmonization. C.Y., R.F.M.H., and S.S.L performed the validation experiments. I.S.C., Y.J, C.Y, and S.S.L. wrote the manuscript.

## Notes

The authors declare no competing financial interest.

ACKNOWLEDGMENT

This work was supported by Hansol RootOne and the Basic Science Research Program through the National Research Foundation of Korea (NRF) (2021R1A3A3002527). S.S.L. acknowledges the Fellowship from the Yangyoung Foundation.

# Supplementary Information

## MoSAiC: Codon Harmonization of Monte Carlo-Based Simulated Annealing for Linked Codons in Heterologous Protein Expression

Yoonho Jeong,[a] Chengcheng Yang,[b] Ryan Fernandez Medina Hariri,[b] Jihoo Kim,[a] Eok Kyun Lee,[a] Younghoon Lee,[a] Won June Kim,[c] Seung Seo Lee,[b] and Insung S. Choi[*a]

[a]Department of Chemistry, KAIST, Daejeon 34141, Korea
[b]School of Chemistry and Chemical Engineering, Highfield Campus, University of Southampton, Southampton SO17 1BJ, United Kingdom
[c]Department of Biology and Chemistry, Changwon National University, Changwon 51140, Korea

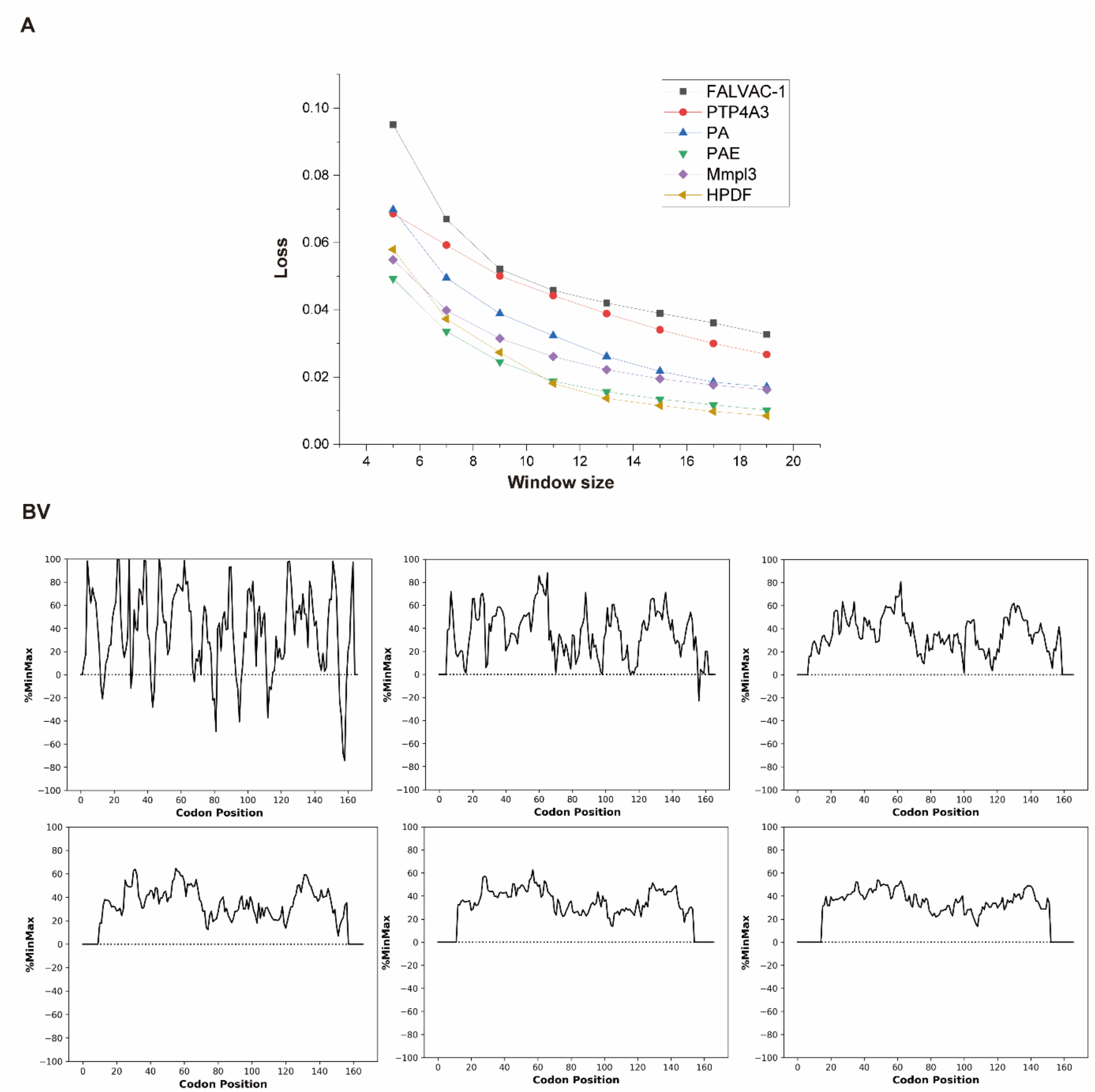

**Figure S1.** Effect of window size on loss and $\%MinMax$ profiles. (A) Normalized loss values for six proteins (FALVAC-1, PTP4A3, PA, PAE, Mmpl3, and HPDF) as a function of window size. (1) For all proteins, the loss decreases monotonically with increasing window size, indicating improved smoothing and agreement with the target profile at larger window sizes. (B) $\%MinMax$ profiles of the RP S10 protein generated using different window sizes, arranged from left to right, top to bottom (window sizes: 5, 10, 15, 20, 25, and 30).

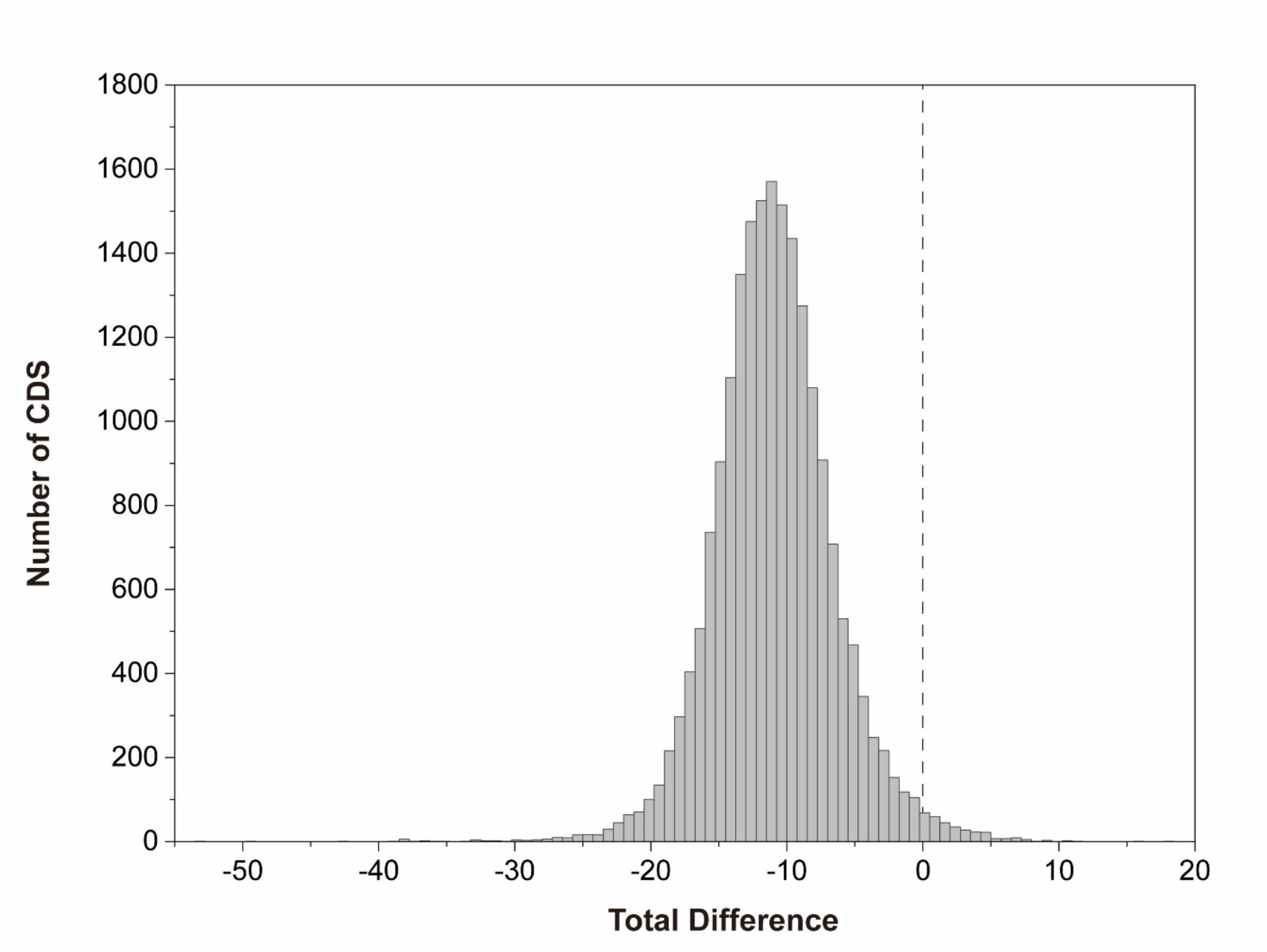

**Figure S2.** Distribution of total $\%MinMax$ profile differences between original and initialized coding sequences. For each sequence, the total difference was calculated as the mean difference across the $\%MinMax$ profiles. CDS: coding sequence.

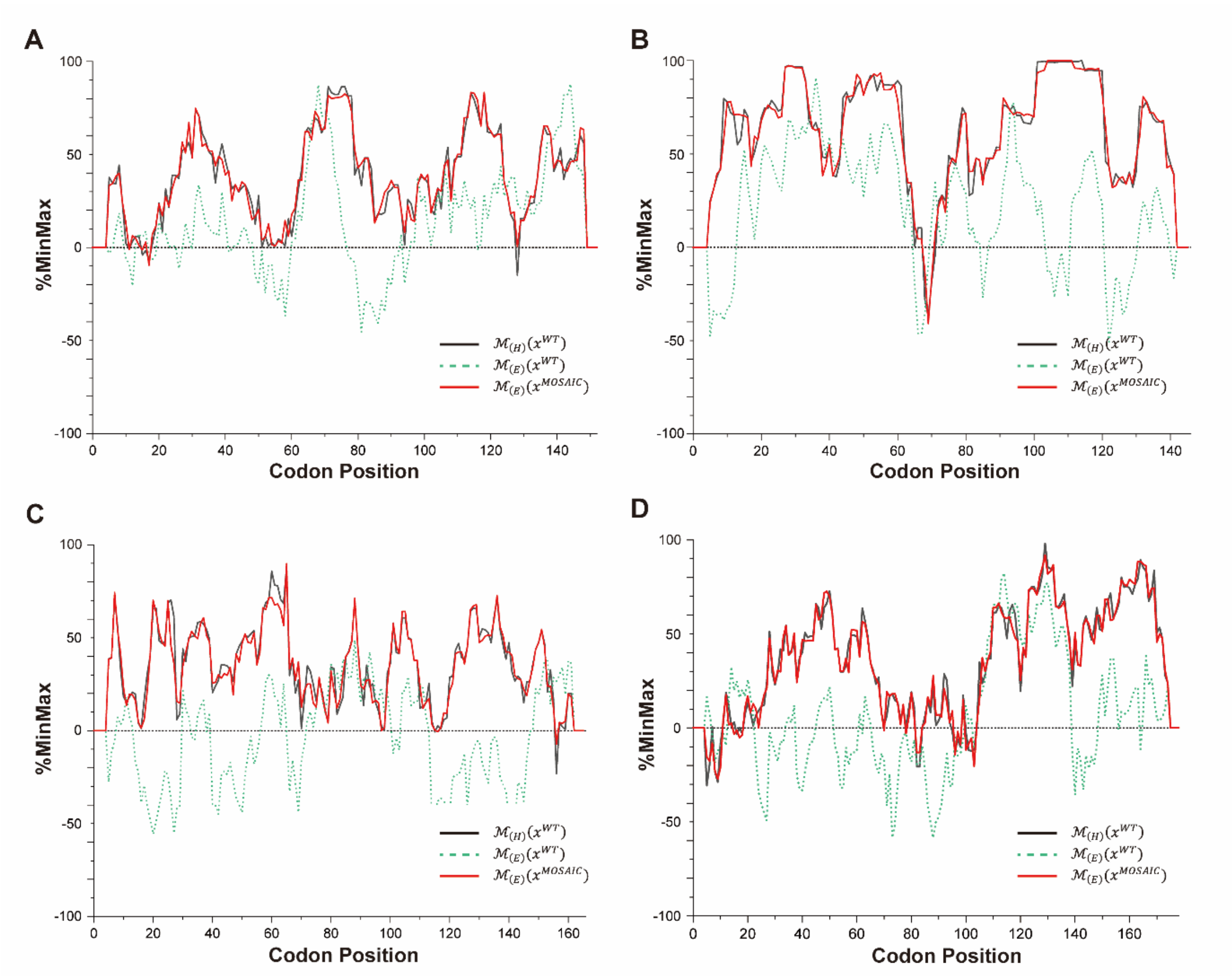

**Figure S3**. %$MinMax$ profiles for (A) RP S18, (B) RP S15, (C) S10, and (D) L11.

**Table S1.** The loss ($\mathcal{L}$) values of the experiments for optimizing $\tau^{(0)}$ values for the proteins. During the experiments, the maximum number of iterations was set to 1,000 times the sequence length.

A. Loss ($\mathcal{L}$) values for RP S18, S15, S10, and L11 from 1 to 17 of initial temperature by 2.

| $\boldsymbol{\tau^{(0)}}$ | S18 | S15 | S10 | L11 |
|---|---|---|---|---|
| 1 | 739.25 | 680.68 | 804.17 | 970.69 |
| 3 | 615.05 | 652.23 | 859.14 | 856.22 |
| 5 | 614.43 | 591.61 | 747.68 | 828.43 |
| 7 | 647.44 | 530.78 | 677.44 | 813.15 |
| 9 | 611.25 | 541.55 | 762.49 | 691.58 |
| 11 | 605.01 | 611.04 | 762.57 | 753.79 |
| 13 | 537.67 | 551.37 | 652.8 | 695.34 |
| 15 | 585.95 | 560.18 | 689.28 | 755.04 |
| 17 | 555.76 | 588.18 | 753.38 | 769.86 |

B. Loss ($\mathcal{L}$) values for RP S18, S15, S10, and L11 from 11 to 15 of initial temperature by 0.5.

| S18 | 11 | 11.5 | 12 | 12.5 | 13 | 13.5 | 14 | 14.5 | 15 |
|---|---|---|---|---|---|---|---|---|---|
| 1 | 589.9 | **563.4** | 581.5 | **554.7** | 632.6 | 577.1 | 624.6 | 566.2 | 566.7 |
| 2 | 626.1 | 671.1 | 554.2 | 661.4 | 562.2 | 584.5 | **551.4** | 598.8 | 587.5 |
| 3 | **568.8** | 620.5 | 588.2 | 633.6 | 631.8 | 602.7 | 567.7 | 599.1 | 588.4 |
| 4 | 644.3 | 587.7 | **552.3** | 616.8 | 618.0 | 662.0 | 565.0 | 589.3 | 623.9 |
| 5 | 579.1 | 605.5 | 601.6 | 626.2 | 611.8 | 650.8 | 554.9 | 653.0 | **547.3** |
| 6 | 574.4 | 636.0 | 618.8 | 559.5 | 634.5 | 620.9 | 598.6 | 625.0 | 618.2 |
| 7 | 595.7 | 630.7 | 613.2 | 577.7 | 543.2 | 536.5 | 687.3 | 645.5 | 633.2 |
| 8 | 592.2 | 664.9 | 617.4 | 581.2 | 593.5 | 608.1 | 538.4 | 598.4 | 611.0 |
| 9 | 632.8 | 584.4 | 586.1 | 648.9 | **533.1** | **532.2** | 616.0 | 598.7 | 547.6 |
| 10 | 586.0 | 612.5 | 553.0 | 575.6 | 584.9 | 573.6 | 564.2 | **552.2** | 570.1 |

| mean | 598.4 | 616.8 | 586.1 | 602.5 | 593.4 | 593.4 | **585.3** | 601.9 | 588.6 |
|---|---|---|---|---|---|---|---|---|---|

| S15 | 11 | 11.5 | 12 | 12.5 | 13 | 13.5 | 14 | 14.5 | 15 |
|---|---|---|---|---|---|---|---|---|---|
| 1 | 554.4 | 585.5 | 565.8 | 585.4 | 533.2 | 595.9 | 580.9 | 545.2 | 538.6 |
| 2 | 540.4 | 574.7 | 578.2 | 562.8 | 556.1 | 608.2 | 531.4 | 529.2 | 564.1 |
| 3 | 579.0 | **515.0** | 570.7 | 556.2 | 541.2 | 577.2 | 562.5 | 634.1 | 551.9 |
| 4 | **539.3** | 581.9 | 563.7 | 555.2 | **522.1** | 572.2 | 556.6 | 552.3 | 538.7 |
| 5 | 618.8 | 551.6 | 559.2 | 554.7 | 525.9 | **521.2** | 609.4 | 605.0 | 512.3 |
| 6 | 584.0 | 564.7 | 549.3 | 565.3 | 558.3 | 544.2 | 590.0 | **523.6** | 585.2 |
| 7 | 540.7 | 546.6 | 539.9 | **529.3** | 536.9 | 562.2 | 541.9 | 564.0 | **513.3** |
| 8 | 564.9 | 559.1 | 562.9 | 569.7 | 532.0 | 564.2 | 526.2 | 556.3 | 536.6 |
| 9 | 553.9 | 534.3 | **514.8** | 569.7 | 556.9 | 598.3 | **522.4** | 553.2 | 577.2 |
| 10 | 556.3 | 531.2 | 532.7 | 544.8 | 531.7 | 536.9 | 542.7 | 586.7 | 556.9 |
| mean | 562.7 | 554.0 | 553.4 | 559.1 | **539.3** | 567.4 | 555.7 | 564.0 | 547.0 |

| S10 | 11 | 11.5 | 12 | 12.5 | 13 | 13.5 | 14 | 14.5 | 15 |
|---|---|---|---|---|---|---|---|---|---|
| 1 | **687.2** | 773.9 | 769.6 | **682.4** | 688 | **707.6** | 692.1 | 714.4 | 749.1 |
| 2 | 736.3 | 750.3 | 725.7 | 720.0 | 765.7 | 727.6 | **687.7** | 721.9 | 708.9 |
| 3 | 693.6 | 698 | 757.1 | 769.7 | 661.7 | 710.0 | 822.3 | 801.5 | 788.9 |
| 4 | 738.6 | 726.3 | 722.8 | 737.1 | 706.1 | 740.7 | 818.4 | 731.0 | 719.3 |
| 5 | 733.8 | **674.9** | 736.4 | 800.7 | 724.5 | 709.3 | 810.7 | 705.1 | 743.5 |
| 6 | 712.8 | 756.1 | **678.9** | 723.6 | 732.7 | 729.6 | 753.5 | 728.8 | 758.3 |
| 7 | 743.0 | 738.6 | 692.1 | 772.3 | 718.7 | 777.8 | 771.7 | 741.4 | 810.7 |
| 8 | 747.7 | 748.9 | 706.1 | 757.4 | 731.1 | 726.1 | 804.0 | **685.6** | 711.5 |
| 9 | 754 | 738.4 | 699.9 | 800.3 | 709.0 | 714.8 | 737.0 | 763.7 | 730.0 |
| 10 | 743.2 | 740.6 | 721.3 | 748.8 | **664.8** | 753.0 | 716.1 | 712.3 | **700.2** |

| mean | 728.7 | 734.1 | 720.5 | 750.4 | **709.6** | 729.3 | 759.7 | 730.0 | 741.3 |
|---|---|---|---|---|---|---|---|---|---|

| L11 | 11 | 11.5 | 12 | 12.5 | 13 | 13.5 | 14 | 14.5 | 15 |
|---|---|---|---|---|---|---|---|---|---|
| 1 | 757.5 | 720.8 | 708.7 | 734.3 | 715.6 | 732.8 | **651.9** | **664.9** | 725.9 |
| 2 | 648.6 | 689.8 | 762.6 | 759.1 | 660.4 | **634.6** | 823.1 | 733.8 | 841.8 |
| 3 | 702.5 | 665.3 | **676.3** | 806.5 | **703.1** | 730.2 | 873.1 | 783.8 | 829.0 |
| 4 | 724.6 | 725.8 | 733.4 | 796.3 | 857.9 | 675.0 | 717.3 | 716.2 | 777.0 |
| 5 | 825.6 | 754.2 | 710.7 | **701.3** | 726.1 | 776.8 | 714.9 | 736.7 | 728.8 |
| 6 | **604.2** | 668.6 | 762.4 | 729.7 | 791.1 | 651.3 | 696.2 | 747.0 | **658.4** |
| 7 | 706.8 | 818.2 | 719.6 | 753.6 | 780.6 | 787.8 | 738.7 | 731.6 | 699.3 |
| 8 | 817.9 | 706.5 | 750.5 | 707.5 | 764.6 | 752.2 | 702.5 | 762.1 | 689.3 |
| 9 | 734.7 | **636.4** | 833.4 | 755.2 | 780.0 | 794.1 | 741.2 | 694.0 | 666.2 |
| 10 | 826.1 | 705.0 | 734.3 | 772.7 | 710.8 | 734.0 | 762.0 | 820.1 | 852.7 |
| mean | 731.3 | **707.4** | 738.1 | 750.9 | 747.1 | 724.9 | 739.7 | 737.9 | 743.6 |

**Table S2.** Loss (*L*) values for RP S18, S15, S10, and L11 at each optimal initial temperature, respectively.

| | S10 (13) | S15 (13) | S18 (14) | L11 (11.5) |
|---|---|---|---|---|
| 1 | 644.4 | 511.1 | 541.3 | 673.1 |
| 2 | 658.6 | 551.1 | 537.9 | 664.0 |
| 3 | 735.6 | 557.6 | **519.2** | 656.4 |
| 4 | **629.9** | 517.4 | 538.4 | 693.1 |
| 5 | 642.5 | **509.6** | 546.5 | 665.6 |
| 6 | 669.2 | 539.8 | 533.4 | 681.7 |
| 7 | 657.1 | 514.4 | 534.0 | 657.7 |
| 8 | 646.4 | 573.2 | 533.7 | 685.1 |
| 9 | 665.0 | 547.4 | 535.3 | 731.0 |

| 10 | 695.2 | 549.5 | 544.9 | **655.0** |
|---|---|---|---|---|
| mean | 663.8 | 536.7 | 536.5 | 675.9 |

**Table S3.** Raw Integrated Density (RID) values obtained by densitometric analysis in ImageJ for each lane and target protein band across four biological replicates. Relative band intensity (% of Total) was calculated by dividing target band raw integrated density by the corresponding construct total lane raw integrated density and multiplying by 100.

| Sample | n | hRPS18 Total Lane RID | hRPS18 Target Band RID | hRPS18 % of Total | wtRPS18 Total Lane RID | wtRPS18 Target Band RID | wtRPS18 % of Total |
|---|---|---|---|---|---|---|---|
| Bacterial Lysate | 1 | 17935067 | 1882680 | 10.497 | 19171985 | 1883456 | 9.824 |
| | 2 | 8421998 | 202715 | 2.407 | 10140485 | 328220 | 3.237 |
| | 3 | 11650719 | 595292 | 5.120 | 9206277 | 367019 | 3.987 |
| | 4 | 7683035 | 657606 | 8.559 | 6816551 | 348620 | 5.114 |
| Soluble Fraction | 1 | 24311730 | 908570 | 3.737 | 25311072 | 885671 | 3.499 |
| | 2 | 10520962 | 467737 | 4.446 | 12216102 | 535807 | 4.386 |
| | 3 | 10198205 | 167165 | 1.639 | 10874234 | 139763 | 1.285 |
| | 4 | 6358727 | 151164 | 2.377 | 5616362 | 68488 | 1.219 |
| Inclusion Body | 1 | 14439175 | 1725443 | 11.950 | 16991527 | 1922803 | 11.316 |
| | 2 | 11465757 | 346950 | 3.026 | 10677389 | 219583 | 2.057 |
| | 3 | 7704073 | 712631 | 9.250 | 10195150 | 184449 | 1.8092 |
| | 4 | 6270024 | 357297 | 5.699 | 5840402 | 180572 | 3.0918 |